\documentclass[a4paper]{jpconf}
\usepackage{graphicx}

\newcommand*{\Ham}{{\cal H}}

\newcommand*{\be}{\begin{equation}}
\newcommand*{\ee}{\end{equation}}
\newcommand*{\bea}{\begin{eqnarray}}
\newcommand*{\eea}{\end{eqnarray}}

\newcommand{\cecoin}{CeCoIn$_5$}
\begin{document}
\title{Tunneling limit of heavy-fermion point contacts}

\author{Mikael Fogelstr\"om$^1$ and Matthias J. Graf$^2$}

\address{$^1$ Department of Microtechnology and Nanoscience, Chalmers, S-412 96 G\"oteborg, Sweden\\
$^2$ Theoretical Division, Los Alamos National Laboratory, Los Alamos, New Mexico 87545, USA}



\begin{abstract}
We present results for a multichannel tunneling model that describes point-contact spectra between a 
metallic tip and a superconducting heavy-fermion system.
We calculate tunneling spectra both in the normal and superconducting state. 
In point-contact and scanning tunneling spectroscopy many heavy-fermion materials, like \cecoin,
exhibit an asymmetric differential conductance,
$dI/dV$, combined with a strongly suppressed Andreev reflection signal in the superconducting state.
For Andreev reflection to occur a junction has to be in the highly transparent limit. Here we focus on the
opposite limit, namely that of low transparency leading to BCS-like $dI/dV$ curves.
We discuss the consequences of a multichannel tunneling model for \cecoin\ 
assuming itinerant electron bands and localized $f$ electrons.
\end{abstract}

\section{Introduction}

Point contact spectroscopy (PCS) and scanning tunneling spectroscopy (STS) have been widely used to study the electronic properties of 
heavy-fermion superconductors (HFS). 
A general problem in this field has been the interpretation of tunneling data, which show asymmetric conductances in the normal state and significant deviations from the standard BTK formalism (Blonder-Tinkham-Klapwijk) \cite{BTK}. In the past, the BTK formalism has been very successful in describing tunneling conductances between metal tips and conventional superconductors, 
while resulting in unphysical parameterizations of the HFS tunneling conductances.

Here we present results for a multichannel tunneling model between a metallic tip and a heavy-fermion superconductor.  These results
are discussed with respect to the anomalous properties observed in \cecoin\
 \cite{goll2005,goll2006,park2008a,park2008b,Ernst2010},
but are readily applied to other heavy fermions (HF) 
like CeCu$_2$Si$_2$, URu$_2$Si$_2$ and many others \cite{ManyMore,Steglich, URS1, URS2}.

\section{Tunneling model}

We model the HF material by two itinerant bands and additional localized surface states,
which may be caused by broken $f$-electron bonds at the surface due to the broken translation symmetry,
\bea
\Ham_{HF}&=&\sum_{\alpha;k,\sigma} {\cal E}_\alpha(k) c^\dagger_{\alpha;k\sigma} c_{\alpha;k\sigma}
+ E_0 \sum_{i\sigma} f^\dagger_{i\sigma} f_{i\sigma} .
\eea
The heavy-fermion Hamiltonian $\Ham_{HF}$ represents two bands of itinerant conduction electrons with band index
$\alpha\in\{{\rm light, heavy}\}$ and localized electrons near the surface with site index $i$.
The operators $c^\dagger_{\alpha;k\sigma}$ ($c_{\alpha;k\sigma}$) create (destroy) an itinerant
electron with momentum $k$ and spin $\sigma$ in band $\alpha$, while operators
$f^\dagger_{i\sigma}$ ($f_{i\sigma}$) create (destroy) an $f$ electron at site $i$ with spin $\sigma$.
${\cal E}_\alpha(k)$ are the respective electronic dispersions and $E_0$ is the energy level
of the localized $f$ electrons.

\begin{figure}[tb]
\centerline{\includegraphics[width=1.0\textwidth,angle=0]{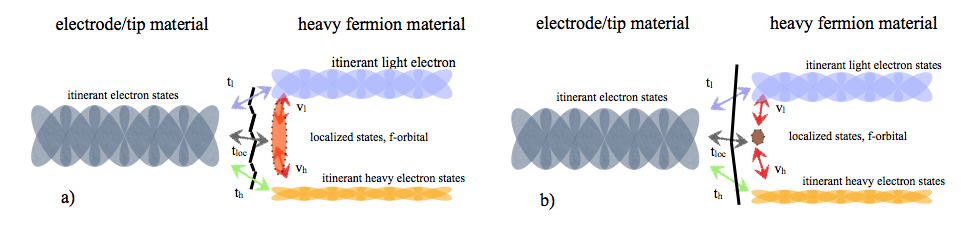}}
\caption{(Color online)
A cartoon of the tunneling processes from the tip of the point contact to the heavy-fermion
material (localized and itinerant electrons), which are necessary to account for the 
measured asymmetry in point-contact junction conductances and reduced Andreev reflection signals. 
In (a) the localized
state appears as a broad resonance at the interface while in (b) the localized state forms a sharp surface state, 
which acts as a resonant tunneling center. 
}
\label{fig:Setup}
\end{figure}

A simple description of a tunneling experiment is comprised of Hamiltonians for the heavy-fermion material, 
the counter electrode, and the transfer or tunneling processes between them:
$\Ham = \Ham_{\rm HF} + \Ham_{\rm electrode} + \Ham_{\rm T}$.
The counter electrode is given by normal conduction electrons
\be
\Ham_{\rm electrode}=\sum_{k,\sigma} {\cal E}_e(k) e^{\dagger}_{k\sigma}e_{k\sigma}  ,
\ee
and the tunneling Hamiltonian describes all possible transfers
\be
\Ham_{\rm T}=\!\! \sum_{\alpha: k,\sigma;k^\prime\sigma^\prime} 
\bigg\lbrack
t^{\alpha}_{k,\sigma;k^\prime\sigma^\prime} e^{\dagger}_{k\sigma} c_{\alpha;k^\prime\sigma^\prime}
+ t^{\alpha}_{k,\sigma;k^\prime\sigma^\prime} c^{\dagger}_{\alpha;k\sigma} e_{k^\prime\sigma^\prime}
\bigg\rbrack
+
\sum_{k,\sigma;\sigma^\prime} 
\bigg\lbrack
t^{loc}_{k,\sigma;\sigma^\prime} e^{\dagger}_{k\sigma} f_{i\sigma^\prime}
+
t^{loc}_{k,\sigma;\sigma^\prime} f^{\dagger}_{i\sigma^\prime} e_{k\sigma}
\bigg\rbrack.
\ee
In addition to the standard overlap integrals $t_\alpha$ between the conduction band in 
the point contact and itinerant heavy-fermion bands there is a finite overlap, 
$t_{loc}$, from the point contact to the localized states in the HF. 
We also account for weak coupling between the localized surface electrons and
itinerant electrons through scattering terms
$v_\alpha$ (see Fig.~\ref{fig:Setup}).
In general, to get a Fano resonance in the conductance one needs interference between different tunneling 
paths \cite{fano1961}.
The resulting differential conductance calculated from this model $\Ham_{\rm T}$ will have an asymmetric Fano line shape. 
Figure \ref{fig:Setup} shows the processes that are active in tunneling  
between a metallic point contact and the HF material. 
Here we extend the picture of conduction through individual quantum channels to a
tunneling model to account for point contacts on a HF material. 
When deriving the general tunneling expression, we consider strong overlap between electron states in the 
contact and the HF compound and thus go
beyond the strict tunneling limit.

We calculate the tunneling current through a quantum channel by employing the standard non-equilibrium Green's function technique
\cite{Schrieffer1963, caroli1971, Cuevas1996, Cuevas2001}.
To further simplify our calculations, we make several assumptions:
(1) The itinerant microscopic Green's functions are described by quasiclassical Green's functions near the Fermi energy. 
(2) It is essential to keep the full energy dependence of the localized Green's function. We assume a single localized level at 
energy $\varepsilon={E}_0$. 
(3) Only the heavy electrons undergo a superconducting transition at $T=T_c$, while the light electrons remain uncondensed. 
For details of this model and formalism see Ref.~\cite{Fogelstrom_preprint}.

\section{Tunneling conductance}

For a multiband tunneling model the differential conductance of a single quantum channel was derived
in Ref.~\cite{Fogelstrom_preprint}. It leads to a generalized Fano expression for the conductance
\bea
\frac{dI}{dV}(V) &=& {\cal D} \frac{e^2}{\hbar} \frac{1}{T} 
\, \int^\infty_{-\infty} \frac {d \varepsilon}{2 \pi} 
\frac{|q \Gamma +\varepsilon-{\tilde E}_0|^2}{\Gamma^2+(\varepsilon-{\tilde E}_0)^2 } \,
\cosh^{-2} \bigg\lbrack\frac{\varepsilon-eV}{2T}\bigg\rbrack .
\label{eq:normalstatecurrent}
\eea
In Eq.~(\ref{eq:normalstatecurrent}) 
${\cal D}$ is the transparency of the junction, ${\tilde E}_0$ is the tunneling-renormalized 
position of the localized energy relative to the Fermi level, $\Gamma$ is the half-width of the resonance,
and $q = q_F + i\, q_{\cal B}$ with $q_{{\cal B}}={\cal B}/\Gamma$.
The conventional Fano parameter $q_F$ controls the resonance shape. 
The additional parameter $q_{\cal B}$
is present for multiband models only, when tunneling through a resonant localized state couples differently
to the HF conduction bands (see below).  The term ${\cal B}$ adds a Lorentzian to the conventional Fano resonance.

For notational convenience, we introduce the following parameterization of the microscopic parameters $t_{\{h,l,loc\}}$ and $v_{\{h,l\}}$
\be
\begin{array}{l}
t_h=\tilde t \sin\alpha \sqrt{\eta_h} \cos\theta_t\\
t_l=\tilde t \sin\alpha \sqrt{\eta_l} \sin\theta_t
\end{array},\,\,
\begin{array}{l}
v_h=\tilde v \sqrt{\eta_0\, \eta_h} \cos\theta_v\\
v_l=\tilde v \sqrt{\eta_0\, \eta_l} \sin\theta_v
\end{array},\,\,
\begin{array}{l}t_{loc}=\tilde t \sqrt{\eta_0} \cos\alpha ,
\end{array}
\ee
where $\tilde t=t\,\sqrt{{\cal{N}}_c\,{\cal{N}}_{HF}}$, $\tilde v=v\, {\cal{N}}_{HF} $ are the effective tunneling elements
with the density of states at the Fermi level ${\cal{N}}_{\{HF,c\}} $ in the heavy fermion (contact) material. 
The factor $\eta_0$ is the fraction of localized surface states and $\eta_{\{h,l\}}$ give
the relative fraction of heavy and light electrons. 
The angles $\theta_t$ and $\theta_v$ 
give the relative overlap integrals between the direct tunneling and the hybridization matrix elements. For
simplicity, we assume $\theta_t=\theta_v\equiv\theta$, which results in ${\cal B}\equiv 0$\,\cite{Fogelstrom_preprint}. 
Finally, the angle $\alpha$ quantifies the relative proportion
of tunneling into a localized state relative to direct tunneling.
The four non-zero phenomenological model parameters introduced in Eq.~(\ref{eq:normalstatecurrent}) now depend on the  microscopic 
parameters $(\tilde t, \tilde v, E_0,\alpha,\theta)$ and the three fractions $\eta_{\{0,h,l\}}$.

\subsection{Tunneling limit in the normal state}

In the tunneling limit, $\tilde t^2\ll1$, we keep the coupling term $\tilde v$ between localized and itinerant states at arbitrary strength
and obtain
\begin{eqnarray}
{\cal D}&\approx &4\eta_i\,\tilde t^2 \sin^2\alpha ,\\
{\tilde E}_0&\approx&{E}_0 -\eta_0 \eta_i \tilde v \tilde t^2\sin 2\alpha ,\\
\Gamma&\approx&\eta_0\lbrack \eta_i\,\tilde v^2+\tilde t^2(\cos^2\alpha-\eta_i^2\tilde v^2\sin^2\alpha)\rbrack ,\\
q_F&\approx&-\frac{\cot\alpha}{\eta_i \tilde v}\lbrack 1-\frac{\tilde t^2}{n \tilde v^2}(\cos^2\alpha+\eta_i^2\tilde v^2 \sin^2\alpha)\rbrack.
\end{eqnarray}
Here $\eta_i=\eta_h\cos^2 \theta +\eta_l\sin^2\theta$ is the itinerant fraction. 
It quantifies the relative weight of direct tunneling into light and heavy bands.
The parameter 
$\tilde v$ is the effective hybridization matrix element at the surface, which is temperature independent. 

In this limit the point-contact spectra measure a $dI/dV$ curve with a Fano line shape and allow us to directly relate
the model parameters to the bulk state of the HF material. At temperatures low compared to the coherence temperature
of the Kondo lattice, $T\ll T_{coh}$,
the HF state can be described as a renormalized Fermi liquid \cite{Hewson1993,lohneysen2007}. 
The density of state factors can be related to the ratio of the bare mass $m$ with the effective mass $m^*$ as
$\eta_h \approx (1 - m/m^*)$  and $\eta_l \approx m/m^*$ \cite{Hewson1993}. 
The temperature dependence of $m/m^*$ can be obtained, for example, from experiment.
For qualitative purposes, we model it as $m/m^* \sim (T/T_{coh})^p$, with $p>0$.
The factor $\eta_0$ describes the fraction of localized states and its physical meaning and value are under debate, 
see for example the discussion by Yang \cite{Yang2008, Yang2009}.

We can gain some physical insight by considering three special scenarios on how the measured peak width $\Gamma$ of the Fano resonance 
should depend on temperature if it was entirely due to the effective mass. 
For simplicity, we keep only leading order  tunneling terms and drop terms proportional to $\tilde t^2$:
\\
{\em (1) The localized states or moments at the interface are not connected to the HF physics of the bulk.} \
In this case, the fraction $\eta_0$, giving the number of surface states, is temperature independent. We obtain
\bea
\Gamma(T) &\approx& \eta_0\,\frac{\tilde v^2}{2} (1+\cos 2\theta (1-2 \frac{m}{\,\,m^*})), 
\\
q_F(T) &\approx& - \frac{ 2\, {\rm cot\,}\alpha }{\tilde v \left[ 1+\cos 2\theta (1-2 \frac{m}{\,\,m^*}) \right] }, 
\eea
where $\Gamma(T)$ decreases with increasing temperature, while $q_F(T)$ increases,
which is inconsistent with PCS measurements for \cecoin, see Ref.~\cite{Fogelstrom_preprint}.
\\
{\em (2) Localized states or moments at the interface originating from the Ce 4f states at energy $E_0$, which are the same as 
in the bulk of \cecoin.} \ 
In this case, it is reasonable to assume that $\eta_0$
is proportional to the number of unscreened moments, i.e., $\eta_0(T)\sim \eta_0\, m/m^*$. We obtain
\bea
\Gamma(T) &\approx& \eta_0\,\frac{m}{\,\,m^*}\frac{\tilde v^2}{2}  (1+\cos 2\theta (1-2 \frac{m}{\,\,m^*})),
\\
q_F(T) &\approx& - \frac{ 2\, {\rm cot\,}\alpha }{\tilde v \left[ 1+\cos 2\theta (1-2 \frac{m}{\,\,m^*}) \right] }, 
\eea
where both $\Gamma(T)$  and $q_F(T)$ increase with $T$ for $T\ll T_{coh}$. The increase
of $\Gamma(T)$ is consistent with the PCS measurements, while that of $q_F(T)$ is not
\cite{Fogelstrom_preprint}.
\\
{\em (3) The dominant temperature behavior of $\Gamma(T)$ is caused by inelastic scattering processes like spin fluctuations.} \
Resistivity experiments on \cecoin\ show that $\rho(T)$ may be described by  a
self-consistent spin-fluctuation theory \cite{bianchi2003}. Similarly, spin-lattice relaxation rates 
$1/T_1 \sim T^{1.3}$ suggest the importance of spin fluctuations \cite{Kawasaki2003, Curro2003, Curro2007}.
In a spin-fluctuation scenario the inelastic broadening 
increases as $\Gamma(T)\sim T^p$, where $p>1$, whereas to leading order
$q_F(T)$ is temperature independent.
Therefore, broadening due to electrons scattering off spin fluctuations is a reasonable interpretation for the 
temperature dependence of $\Gamma(T)$ found in \cecoin\ \cite{Fogelstrom_preprint}
and more recently in URu$_2$Si$_2$ \cite{URS1, URS2}.
We conclude that the self-consistent incorporation of inelastic scattering processes into our calculations,
similar to Ref.~\cite{lofwander2005},
is an important next step to account for the observed line broadening.
   
\subsection{Tunneling limit in the superconducting state}

\begin{figure}[t]
\centerline{\includegraphics[width=1.0\textwidth,angle=0]{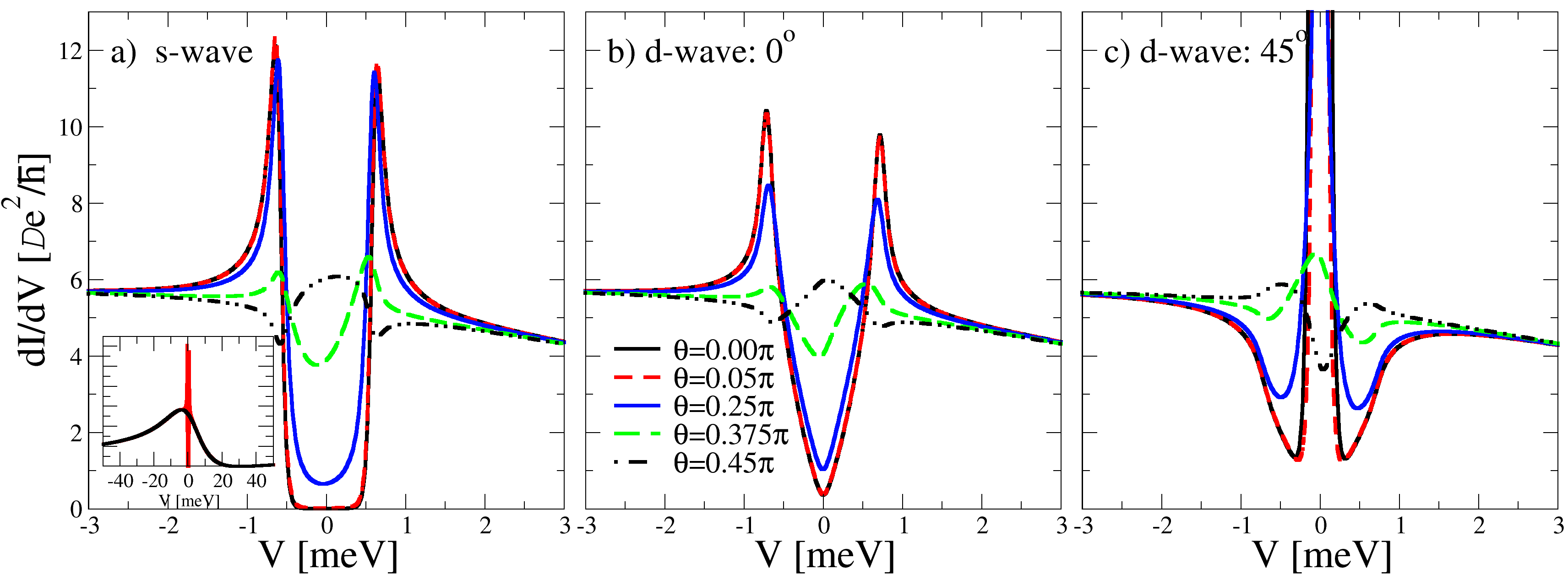}}
\caption{(Color online)
Conductances for an HFS in the tunneling limit.  The parameters of the curves are ${\cal {D}}=0.005, E_0=2.01\, {\rm meV}, \Gamma=13.0 \,{\rm meV}, q_F=-2.16, \alpha=\pi/20$, and $T=0.41$ K with $T_c=2.3$ K.
In panels (a)-(c) we vary the heavy-to-light tunneling parameter $\theta$. 
The conductances are calculated both for an s-wave and d-wave superconducting state, setting $\Delta=0.6 \,{\rm meV}$. As shown, the superconducting $dI/dV$ is very sensitive to the co-tunneling into two bands.
The inset in panel (a) shows the Fano shaped conductance in the normal state at $T>T_c$ with a 
superimposed conductance curve computed at $T<T_c$.
}
\label{fig:dIdV}
\end{figure}

It is necessary to study tunneling in the superconducting state, where Andreev reflection processes modify the conductance, because one can extract
additional information about the microscopic origin of the Fano conductance. 
The unknown relative ratio of tunneling into heavy or light bands can be determined as the $dI/dV$ depends in a non-linear way on $\theta$. At very low temperatures, $T<T_c\ll T_{coh}$, one can assume that $m/m^*$ has saturated to its low-$T$ limit $\sim 1-\eta_{HF}$. 
Thus we write for $\eta_i$ 
defined in the previous section $\eta_i \sim \eta_{HF} \cos^2\theta+(1-\eta_{HF}) \sin^2\theta$
and study the superconducting conductance as a function of $\theta$ and fixing the numerical value of $\eta_{HF}$ to $0.9$.
For simplicity, we assume that only heavy electrons become superconducting, while light electrons remain uncondensed.
In our model, we consider both an s-wave and d-wave superconducting order parameter. For the d-wave case we show results for  two principal orientations. The orientation of the HF crystal lattice 
relative to the interface normal is either antinodal ($0^o$) or nodal  ($45^o$) leading to strikingly different tunneling conductances important for order parameter spectroscopy of the nodes.

In Fig.~2 we plot the conductances in the superconducting state calculated in the tunneling limit by setting ${\cal {D}}=0.005$.
The Fano parameters are chosen to fit experimentally observed $dI/dV$ curves just above
$T_c=2.3$ K in \cecoin. Hence we set $E_0=2.01 \,{\rm meV}$, $\Gamma=13.0 \,{\rm meV}$, and $q_F=-2.16$ in agreement with PCS measurements
with a Au-tip on \cecoin, see Fig.~3 in Ref.~\cite{Fogelstrom_preprint}.
Since the $dI/dV$ curve is only weakly dependent on the tunneling angle $\alpha$ in the superconducting state, we set in our calculations $\alpha=\pi/20$, i.e., stronger tunneling through the localized state compared to the itinerant bands. 
In the case of the ideal one-band superconducting tunneling model, i.e. $\theta=0$, we recover the known $dI/dV$ for all three cases considered. 
In Fig.~2(a) we see the usual BCS superconducting density of states on top of a Fano
background. In the inset of panel (a) we show the voltage dependence of the $dI/dV$ over a large voltage range. In panels (b) and (c) the case of
$\theta=0$ reproduces the $V$-shaped $dI/dV$ for the $0^o$-junction (panel b), 
while for the $45^o$-junction we see 
the hallmark zero-bias peak (panel c). 
Note that for tunneling into the heavy band only, $\theta=0$, the condensate suppresses the normal state $dI/dV$ by 100\% for voltages below the gap, $-\Delta < {\rm eV} < \Delta$.  

Incorporating the additional possibility of tunneling into a band of uncondensed light electrons has two main effects on the $dI/dV$ characteristics.
First, we see that the coherence peaks in the $dI/dV$ curves are reduced from a 100\% effect for small $\theta$
to a $\sim $10\%-effect, when $\theta > \pi/4$. 
Second, the $dI/dV$ curves are qualitatively changed in shape from their ideal one-band $\theta=0$ appearance. In panel (a) we see
that the sub-gap conductance below $\Delta$ can even be increased above its normal-state value, resembling a high-transmission
sub-gap conductance but with a strongly suppressed Andreev reflection signal.
Next, turning to the d-wave conductances we see that in general, for a fixed set of microscopic parameters,
there is a strong dependence on the junction orientation. This property may be used as a smoking gun to identify a d-wave symmetry. However, if we assume that PCS junctions are made on surfaces with different crystal orientations and that these PCS junctions
have very different  characteristics with different sets of microscopic parameters, 
then examining the curves in panels (b) and (c) may not be as distinctive as expected.
In this situation one can find cases where the zero-bias peak is strongly suppressed making the distinction 
between principal nodal/antinodal orientations less striking.
In fact, it may even be difficult to convincingly discriminate between
an s-wave and d-wave superconducting state based on a few tunneling curves alone.

\section{Conclusions}

We presented results for  a  multichannel tunneling model for a superconducting
heavy-fermion material. On the large voltage scale the derived conductances
have the features of Fano-like  $dI/dV$ characteristics. 
This allows us to extract the microscopic model parameters that
describe the relevant tunneling processes. Assuming a modified Fermi-liquid state at low temperatures and a tunneling contact in the tunneling limit, i.e., low transparency,
we find for the normal state of \cecoin\ that the extracted Fano parameters are most likely describing localized surface states and not directly probing the formation of the heavy-fermion state and the temperature behavior of the effective mass. 
In the superconducting state, the calculated $dI/dV$ curves demonstrate that additional microscopic
information can be obtained about tunneling into heavy vs.\ light bands, which is not possible 
from studying the normal state alone.
Finally, we investigated how the $dI/dV$ characteristics depend on the pairing symmetry of the superconducting state. We found that in order to
reliably extract information from tunneling experiments
about the symmetry of the superconducting order parameter detailed modeling of the tunneling processes
is required, which goes beyond a conventional BTK analysis. 

\subsection*{Acknowledgments}
We benefited from discussions with T. L\"ofwander, W. K. Park. L. H. Greene, G. Goll, A. V. Balatsky, Y. Dubi and P. W\"olfle.
M. F.\ was supported by the Swedish Research Council.
M. J. G.\ was  supported in parts by the 
U.S.\ DOE at Los Alamos National Laboratory
under contract No.~DE-AC52-06NA25396 and the Office of Science for BES.

\end{document}